\documentclass[twocolumn,superscriptaddress,prl]{revtex4}
\usepackage{graphicx}
\usepackage{amsmath}
\usepackage{amssymb}
\usepackage{bm}
\begin{document}
\title{Critical Current \boldmath{$0$-$\pi$} Transition in Designed Josephson Quantum Dot Junctions}
\author{H. Ingerslev J\o rgensen}
\email{hij@fys.ku.dk} \affiliation{Nano-Science Center, Niels Bohr
Institute, University of Copenhagen, Universitetsparken 5,
DK-2100~Copenhagen \O , Denmark}
\author{T. Novotn\' y}
\affiliation{Department of Condensed Matter Physics, Faculty of
Mathematics and Physics, Charles University, Ke Karlovu 5, 121 16
Prague, Czech Republic}\affiliation{Nano-Science Center, Niels Bohr
Institute, University of Copenhagen, Universitetsparken 5,
DK-2100~Copenhagen \O , Denmark}
\author{K. Grove-Rasmussen}
\affiliation{Nano-Science Center, Niels Bohr Institute, University
of Copenhagen, Universitetsparken 5, DK-2100~Copenhagen \O ,
Denmark}
\author{K. Flensberg}
\affiliation{Nano-Science Center, Niels Bohr Institute, University
of Copenhagen, Universitetsparken 5, DK-2100~Copenhagen \O ,
Denmark}
\author{P. E. Lindelof}
\affiliation{Nano-Science Center, Niels Bohr Institute, University
of Copenhagen, Universitetsparken 5, DK-2100~Copenhagen \O ,
Denmark}
\date{\today}
%
%
%
\begin{abstract}
We report on quantum dot based Josephson junctions designed
specifically for measuring the supercurrent. From high-accuracy
fitting of the current-voltage characteristics we determine the full
magnitude of the supercurrent (critical current). Strong gate
modulation of the critical current is observed through several
consecutive Coulomb blockade oscillations. The critical current
crosses zero close to, but not at, resonance due to the so-called
$0$-$\pi$ transition in agreement with a simple theoretical model.
\end{abstract}
\maketitle
\begin{figure*}
\centerline{\includegraphics[width=0.6\textwidth]{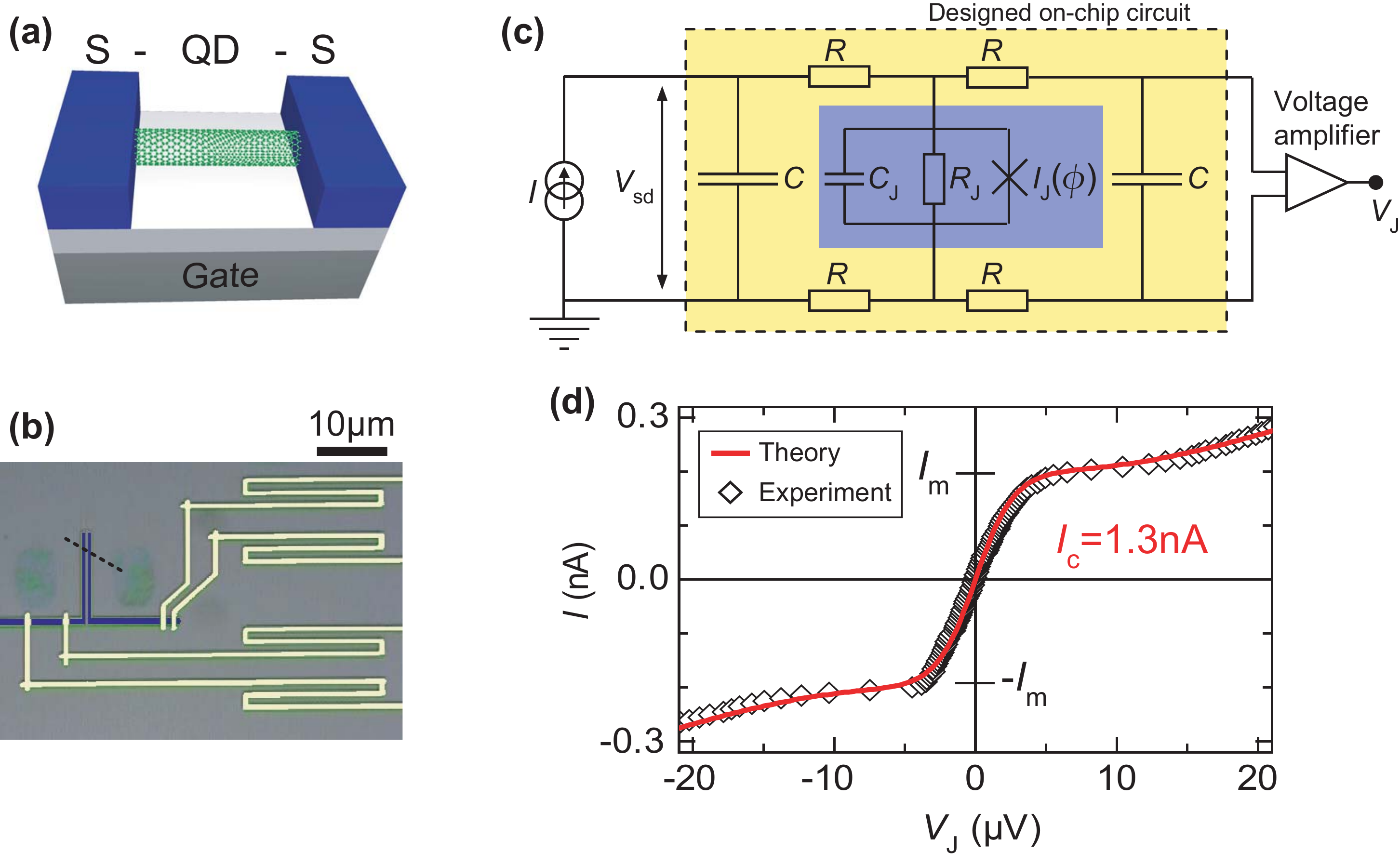}}
\caption{\textbf{Sample design and measurement scheme.}
(a) Schematic illustration of the Superconductor(S)-Quantum
dot(QD)-Superconductor(S) part of the device, made of a single wall
carbon nanotube and a superconducting trilayer of 5\,nm Ti, 60\,nm
Al, and 5\,nm Ti, with transition temperature $T_{\rm c}=0.75$\,K
and energy gap $\Delta = 0.1$\,meV \cite{hijkgr}.
(b) Colored optical image of the device. Blue leads are the
superconducting electrodes with a spacing of 300\,nm, contacting a
nanotube (represented by a dashed line). Yellow leads are long thin
Cr/Au (15\,nm/15\,nm) leads connecting the superconducting leads to
large-area contact pads, outside the image.
(c) Schematic circuit diagram of the device with four-probe
current-controlled measurement setup. Blue square: S-QD-S junction
as depicted in (a). Yellow region: The resistors $R$ are the yellow
leads in (b), and the capacitors $C$ are large-area contact pads.
(d) Current ($I$) versus junction voltage ($V_{\rm J}$) close to a
resonance (see the corresponding arrow in Fig.~\ref{fig:fit}(b)).
Measurement (diamonds) are fitted with Eq.~\eqref{eq:IZ} (red line)
yielding a critical current $I_{\rm c}=1.3$\,nA and a junction
resistance $R_{\rm J}=90$\,k$\Omega$.} \label{fig:setup}
\end{figure*}

%
%
%
%
Formation of a quantum dot (QD) between two superconductors, called
a Josephson QD junction, enables study of supercurrent through a
single energy level of an artificial atom (the QD).
Supercurrent \cite{kasumov,yong-joo,takesue} through a QD
\cite{pablo,hijkgr,wolfgang,kgrhij,jordan} is a dissipationless
current which flows only when a constant phase difference across the
QD is maintained \cite{steinbach,joyez}.
Uncontrolled fluctuations in the phase has so far prevented
observation of the full magnitude of the supercurrent (the critical
current).
Here we analyze Josephson QD junctions, created in a single wall
carbon nanotube, with a carefully designed on-chip circuit that
controls phase fluctuations and thereby allows extraction of the
critical current.
The analysis reveals a reversal of the critical current for every
electron added to the QD, due to the so-called $0$-$\pi$ transition
\cite{ryazanov,baselmans,shi-ptp-69,glaz-jetpl-89,cle-prb-00,spi-prb-91,vecino,and-prb-95,roz-prb-01},
in agreement with a simple theoretical model.
The realization of a well controlled nano-scale Josephson junction
opens new routes toward superconducting quantum bits
\cite{nakamura,vion,chiorescu}.
\newline\indent
When a Josephson QD junction is placed in an electromagnetic
environment the phase difference across the QD ($\phi$) becomes a
dynamical variable, which determines the time-averaged current, $I$,
and voltage drop, $V_{\rm J}=\langle(\hbar/2e)\,d\phi/dt\rangle$,
across it. The $I$-$V_{\rm J}$ characteristics of the junction thus
depends in a non-linear fashion on the phase dynamics induced by the
electromagnetic environment. Inspired by Refs.~\cite{joyez,pablo} we
utilize a designed external circuit in order to control the phase
fluctuations which enables us to infer the true magnitude of the
critical current, $I_{\rm c}$, from the measurable critical
current/switching current, $I_{\rm m}$, by a theoretical fitting
procedure. $I_{\rm m}$ can significantly differ from $I_{\rm c}$ as
demonstrated previously for single wall carbon nanotube (SWCNT)
based Josephson junctions \cite{pablo,hijkgr,kgrhij}.
\newline\indent
To be able to design the external circuit we model our sample
(Fig.~\ref{fig:setup}(a) and (b)) by an extended resistive and
capacitively shunted junction (RCSJ) model \cite{pablo}, yielding
the schematic circuit diagram shown in Fig.~\ref{fig:setup}(c).
The real Josephson QD junction is represented by an ideal Josephson
junction with current-phase relation $I_{\rm J}(\phi)$, in parallel
with a junction capacitor $C_{\rm J}$, and junction resistor $R_{\rm
J}$ accounting for current carried by multiple Andreev reflections.
The resistor $R_{\rm J}$ generally depends on $V_{\rm J}$ and gate
voltage ($V_{\rm gate}$) but for small enough voltage as used in our
analysis $V_{\rm J}\ll\Delta/e$, where $\Delta \sim 0.1$\,meV is the
superconducting energy gap, we approximate $R_{\rm J}$ to depend
only on $V_{\rm gate}$. Assuming a sinusoidal current-phase relation
($I_{\rm J}(\phi)=I_{\rm c}\sin\phi$) the dynamics of $\phi$ in the
circuit of Fig.~\ref{fig:setup}(c) becomes equivalent to the damped
motion of a fictitious particle in the so-called tilted washboard
potential \cite{tinkhambog}. The damping of the motion of this
particle is characterized by the quality factor \cite{joyez,pablo}
\begin{equation}\label{eq:q}
Q = \frac{\sqrt{\hbar[C(1+R/R_{\rm J})+C_{\rm J}]/(2eI_{\rm
c})}}{RC+\hbar/(2eI_{\rm c} R_{\rm J})}
\end{equation}
given by the ratio of the local minimum oscillation frequency to the
friction coefficient. Low $Q$ implies higher probability of trapping
the particle in a potential minimum resulting in a constant phase
difference and, thus, the observation of the supercurrent.
\newline\indent
We have used Eq.~\eqref{eq:q} to design overdamped samples shown in
Fig.~\ref{fig:setup}(b). The superconducting electrodes (blue leads
in Fig.~\ref{fig:setup}(b)) are made with a small area to reduce the
junction capacitance, $C_{\rm J}\sim 1$\,fF. Yellow leads,
connecting the superconducting electrodes to contact pads, are long
thin normal metal wires designed to have a large resistance,
measured to be $R\sim 1.5$\,k$\Omega$. The contact pads are
fabricated with a large area to increase their capacitance, $C \sim
2$\,pF. By inserting the above mentioned values and $R_{\rm J} \geq
h/e^2$ (see Fig.~\ref{fig:spectroscopy}(a)), into Eq.~\eqref{eq:q}
we get a strong damping $Q<1$ for $I_{\rm c}\geq 0.1$nA which is
also the smallest critical current we have been able to measure (see
Fig.~\ref{fig:fit}(b)).
\begin{figure}
\begin{center}
\includegraphics[width=0.48\textwidth]{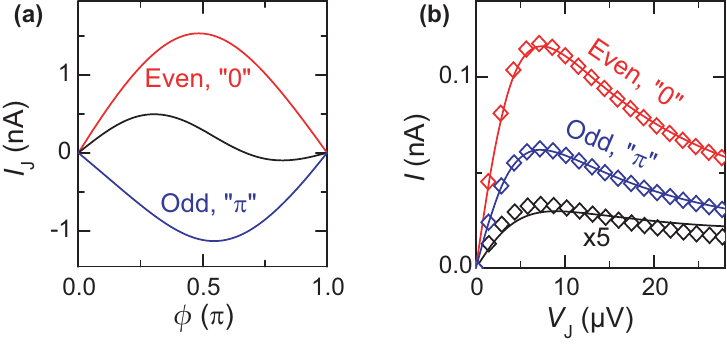}
\end{center}
\caption{\textbf{Fitting procedure for extracting the critical
current.}
(a) Theoretical current-phase relations calculated using a
phenomenological ``external exchange" model \cite{vecino}, with
$E_{\rm ex}\equiv U/2=15\Delta,\,\Gamma=\Gamma_{\rm L}+\Gamma_{\rm
R}=11\Delta,\,\Gamma_{\rm L}=\Gamma_{\rm
R},\,T=\Delta/10,\,I_0=e\Delta/\hbar\simeq 25$\,nA. The three curves
are calculated very close to resonance at $\Delta \varepsilon =
U/30, -U/60, -U/15$ from top to bottom (see arrows in
Fig.~\ref{fig:fit}(a)), where $\Delta \varepsilon$ is the potential
on the quantum dot measured from resonance. Even though this model
is phenomenological it seems to capture correctly all qualitative
features of the $0$-$\pi$ transition known from the numerical
renormalization group \cite{choi} or quantum Monte Carlo
\cite{siano} calculations.
Diamonds in (b) are $I$-$V_{\rm J}$ curves calculated using the
current-phase relations from (a) in the full theory \cite{amb-hal},
valid for a general current-phase relation and in the presence of
thermal fluctuations. Eq.~\eqref{eq:IZ} is fitted to the diamonds
(solid lines) yielding very good fits in the whole gate voltage
range apart from the closest vicinity of the $0$-$\pi$ transitions,
which is below the experimental resolution. (See also the Supporting
Information, Section~S2).} \label{fig:theory}
\end{figure}
\begin{figure}
\begin{center}
\includegraphics[width=0.48\textwidth]{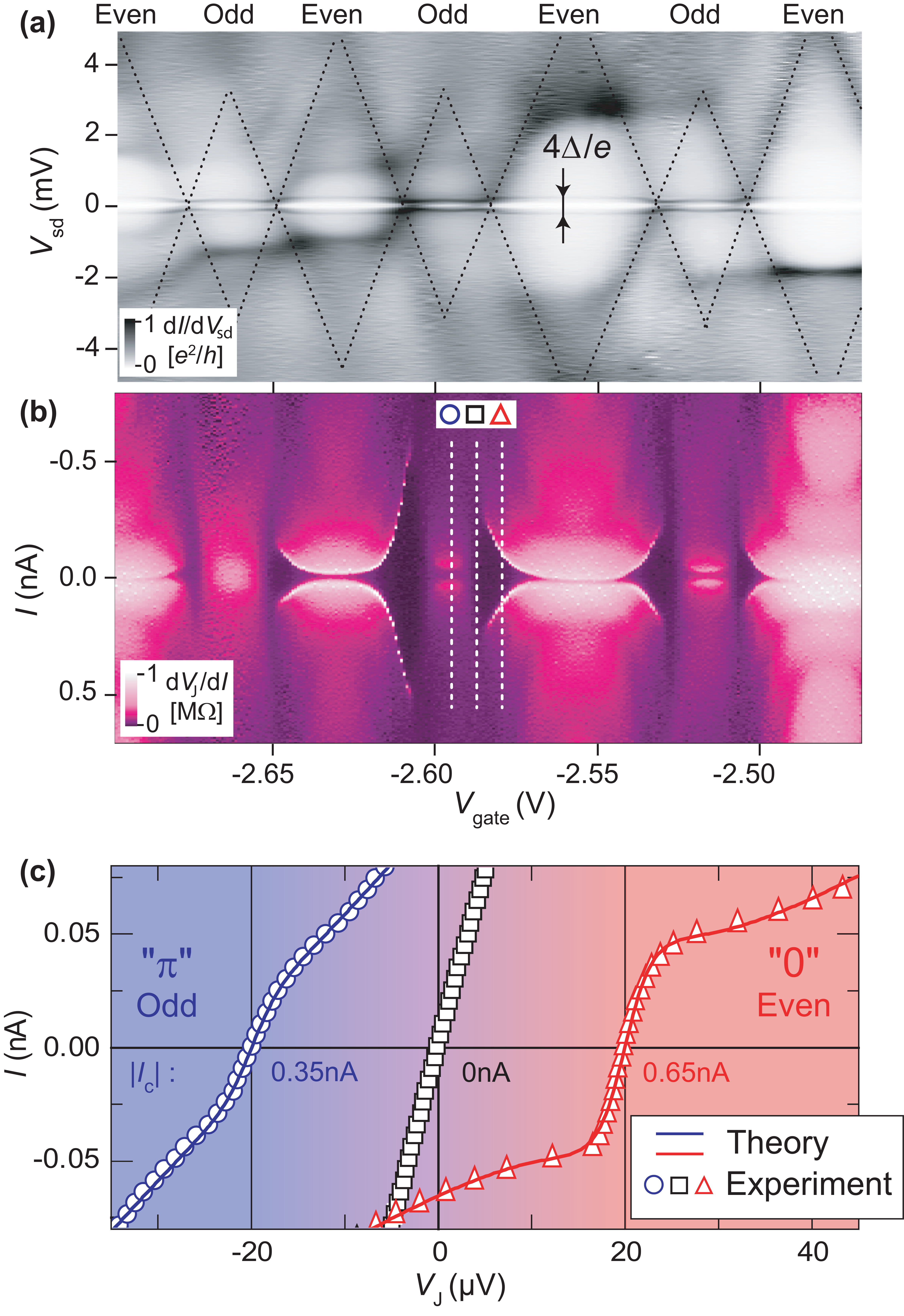}
\end{center}
\caption{\textbf{Supercurrent versus odd and even electron
occupation.} (a) Surface plot of the differential conductance
(d$I$/d$V_{\rm sd}$) versus bias voltage ($V_{\rm sd}$) and gate
voltage ($V_{\rm gate}$). Coulomb blockade diamonds, indicated with
black dotted lines, are alternating in size between large and small
with a corresponding even and odd number of electrons localized on
the QD. (b) Surface plot of differential resistance (d$V_{\rm
J}$/d$I$) versus applied current ($I$) and $V_{\rm gate}$, in the
same gate-voltage range as in (a). (c) Three $I$-$V_{\rm J}$ curves
from (b) at indicated positions, the right and left graphs are
shifted by 20\,$\mu V$ for clarity. Circles are measured with odd
occupation on the dot in the $\pi$ junction regime, squares at the
$0$-$\pi$ transition point, and triangles with even occupation on
the dot in the $0$ junction regime. The solid lines are fits using
Eq.~\eqref{eq:IZ} yielding critical currents of 0.65\,nA(0.35\,nA)
for even(odd) electron occupation, and zero at the transition
point.} \label{fig:spectroscopy}
\end{figure}
\begin{figure}[b]
\begin{center}
\includegraphics[width=0.48\textwidth]{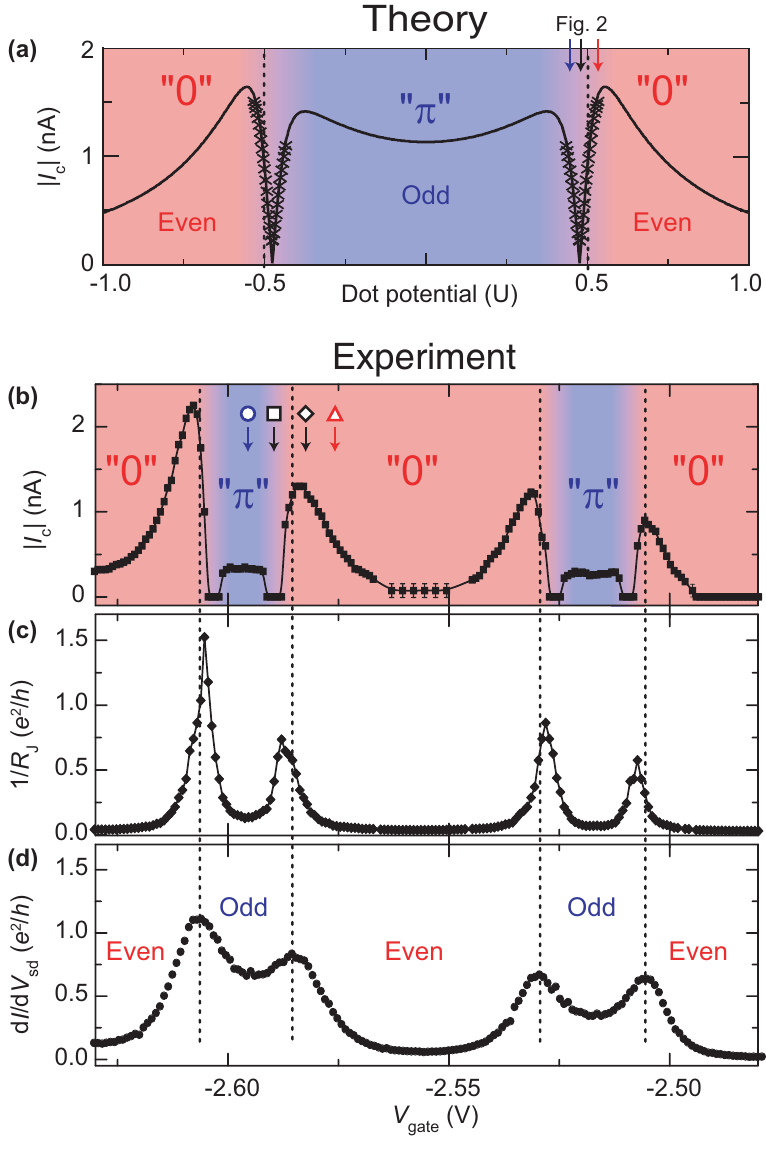}
\end{center}
\caption{\textbf{Critical current \boldmath{$0$-$\pi$} transition
compared to theory.}
(a) Solid line is $|I_{\rm c}|$ versus potential on the quantum dot
obtained as $|I_{\rm J}(\pi/2)|$ from \cite{vecino} with the same
parameters as in Fig.~\ref{fig:theory}. Crosses are the critical
currents obtained by fitting Eq.~\eqref{eq:IZ} to the full theory as
described in Fig.~\ref{fig:theory}(b). Apart from very close to the
transition point, the match between the two approaches is very good.
(b) and (c) Experimental critical current $|I_{\rm c}|$ and junction
conductance $1/R_{\rm J}$ as extracted from fitting measured
$I$-$V_{\rm J}$ curves to Eq.~\eqref{eq:IZ}, versus the gate voltage
$V_{\rm gate}$. The error-bars in (b) at the plateau around $V_{\rm
gate}=-2.56$\,V show the estimated precision of the fit.
(d) Normal state differential conductance ($B = 150$\,mT) at zero
bias, in the same gate-voltage range as in (b) and (c). The dotted
vertical lines indicate the positions of charge-degeneracy
resonances.
Note the good correspondence between the phenomenological model in
(a) and the measurement in (b), both the observed magnitude of the
critical current as well as the shift toward odd occupation of all
the $0$-$\pi$ transition points are in agreement with the theory.}
\label{fig:fit}
\end{figure}
\newline\indent
The measurements are performed in a $^3$He-$^4$He dilution fridge at
a base electron temperature of 75\,mK. In Fig.~\ref{fig:setup}(d) we
show a current biased four-probe measurement of an $I$-$V_{\rm J}$
curve at very low bias voltage $V_{\rm sd} \ll \Delta/e$. In this
measurement, $I_{\rm m}$ is the point where the curve has a large
change in slope (as indicated in Fig.~\ref{fig:setup}(d)), giving
$I_{\rm m} \sim 0.2$\,nA. For $|I|<I_{\rm m} (< I_{\rm c})$ the
particle in the tilted washboard potential has a high probability
after thermal activation out of one potential minimum to be
subsequently retrapped in the next potential minimum. The motion of
the particle is therefore diffusive, leading to a small average
phase velocity, i.e., a low but finite junction voltage. The branch
in the $I$-$V_{\rm J}$ curve at $|I| < I_{\rm m}$, called the
diffusive supercurrent branch, therefore has finite resistance. As
the current is ramped further up $I>I_{\rm m}$ the friction in the
tilted washboard potential is no longer large enough to retrap the
particle once it is activated out of a minimum. This leads to a high
phase velocity and hence the supercurrent is averaged out.
\newline\indent
In order to find $I_{\rm c}$, we fit the measured $I$-$V_{\rm J}$
curves to an overdamped ($Q<1$) extended RCSJ model. For the
overdamped Josephson junction we need to consider the classical
dynamics only. In Ref.~\cite{ivan-zil} the overdamped RCSJ model
(without $R_{\rm J}$) was calculated for sinusoidal current-phase
relation and was extended to general current-phase relations in
Ref.~\cite{amb-hal}. High-resistance tunnel junctions have
sinusoidal current-phase relations, but QD junctions with
resistances comparable to the resistance quantum $h/e^2$ may have
non-sinusoidal current-phase relation (see, e.g., Eq.~(2) in
Ref.~\cite{hijkgr}). Nevertheless, we approximate the current-phase
relation by the simple sinusoidal form parameterized by $I_{\rm c}$,
and we justify this approximation in Fig.~\ref{fig:theory} by
comparing with a theoretical calculation where the full
non-sinusoidal current-phase relation is included (see also the
Supporting Information, Section~S2). We generalize the theory of
Ref.~\cite{ivan-zil} to include current carried via multiple Andreev
reflections by the resistor $R_{\rm J}$ which is assumed to be much
larger than the lead resistance $R$. The subsequent fitted values of
$R_{\rm J}$ are consistent with this assumption, see
Fig.~\ref{fig:fit}(c). Under these assumptions ($I_{\rm
J}(\phi)=I_{\rm c}\sin\phi$ and $R_{\rm J}\gg R$) the $I$-$V_{\rm
J}$ curve is given parametrically via $V_{\rm sd}$ by the relations
\begin{equation}\label{eq:IZ}
\begin{split}
&I(V_{\rm sd})=I_{\rm c}\, \textrm{Im}\!\! \left[ \frac{I_{1 - i
\eta (V_{\rm sd})}(I_{\rm c}\hbar/2ek_B T)}{I_{-i \eta (V_{\rm
sd})}(I_{\rm c}\hbar/2ek_B T)}
\right] + \frac{V_{\rm J}(V_{\rm sd})}{R_{\rm J}},\\
&V_{\rm J}(V_{\rm sd})=V_{\rm sd}-RI(V_{\rm sd}),
\end{split}
\end{equation}
where $I_{\alpha}(x)$ is the modified Bessel function of the complex
order $\alpha$, and $\eta (V_{\rm sd}) = \hbar V_{\rm sd}/2eRk_B T$.
Since all the parameters entering Eq.~\eqref{eq:IZ} apart from
$I_{\rm c}$ and $R_{\rm J}$ are experimentally known we can use
$I$-$V_{\rm J}$ curves to determine both $I_{\rm c}$ and $R_{\rm
J}$. The solid curve in Fig.~\ref{fig:setup}(c) is a fit of
Eq.~\eqref{eq:IZ} to the measured data, yielding $I_{\rm c}=1.3$\,nA
and $R_{\rm J}=90$\,k$\Omega$. The theory seems to capture the
experimental measurement very well, which shows that the suppression
of $I_{\rm c}=1.3$\,nA into the diffusive supercurrent branch with
$I_{\rm m} = 0.2$\,nA indeed is caused by thermal fluctuations. A
total of five samples were fabricated with different contact
resistances. They all showed $I$-$V_{\rm J}$ curves with
qualitatively the same behavior as the sample in
Fig.~\ref{fig:setup}(c) and showed excellent fits to
Eq.~\eqref{eq:IZ} (see Supporting Information, Section~S1).
\newline\indent
In Fig.~\ref{fig:spectroscopy}(a) we show a color-scale plot of the
differential conductance (d$I$/d$V_{\rm sd}$) versus $V_{\rm sd}$
and $V_{\rm gate}$. Coulomb blockade diamonds, indicated with black
dotted lines, alternate in size between large and small as $V_{\rm
gate}$ is increased. This indicates that the QD has two-fold spin
degeneracy of each discrete energy level, with an odd (even) number
of electrons on the QD in the small (large) diamond. Coulomb
repulsion energy $U \sim 3$\,meV and level spacing $\Delta E \sim
2.5$\,meV are extracted from this plot. The edges of the Coulomb
diamonds are somewhat blurred due to the tunnel coupling between QD
and leads. The tunnel coupling can vary from device to device and in
the Supporting Information, Section~S1, we show measurements on
another device (fabricated the same way) exhibiting sharer edges of
the diamonds due to a slightly lower tunnel coupling. At low bias
voltage two parallel conductance ridges (indicated with black
arrows) are seen, reflecting the peak in density of states of the
superconductors at $V_{\rm sd}=\pm 2\Delta/e$, yielding $\Delta \sim
0.1$\,meV. The energy broadening $\Gamma=\Gamma_{\rm L}+\Gamma_{\rm
R}$, where $\Gamma_{\rm L (R)}$ is the coupling to the left (right)
lead, and the asymmetry $\Gamma_{\rm L}/\Gamma_{\rm R}$ of each
Coulomb oscillation are determined in the normal state ($B=150$\,mT)
by fitting the even valley part of the Coulomb oscillation peaks,
Fig.~4(d), to a Lorentzian (see Supporting Information, Section~S3).
This gives approximately constant values of $\Gamma \sim 1.1\,$meV
and asymmetry parameters $\Gamma_L/\Gamma_R$ in the range 1 to 4.
These values are used in the theoretical plots in
Fig.~\ref{fig:theory} and Fig.~\ref{fig:fit}(a).
\newline\indent
In Fig.~\ref{fig:spectroscopy}(b) we plot the differential
resistance d$V_{\rm J}$/d$I$ as a function of $I$ and $V_{\rm
gate}$. Three representative $I$-$V_{\rm J}$ curves selected from
Fig.~\ref{fig:spectroscopy}(b) at indicated positions are shown in
Fig.~\ref{fig:spectroscopy}(c). The diffusive supercurrent branch is
observed at most gate-voltages also at the resonances, but not in a
narrow gate-region close to each resonance (open square in
Fig.~\ref{fig:spectroscopy}(c) and Fig.~\ref{fig:fit}(b)). Full
gate-voltage dependence of $I_{\rm c}$, and $R_{\rm J}$ obtained by
fitting the $I$-$V_{\rm J}$ curves from
Fig.~\ref{fig:spectroscopy}(b) to Eq.~\eqref{eq:IZ} is shown in
Fig.~\ref{fig:fit}(b) and (c), together with the normal-state
($B=150$\,mT) zero-bias conductance in Fig.~\ref{fig:fit}(d). We
observe that $I_{\rm c}$ oscillates in accordance with the number of
electrons (odd or even) localized on the QD. $I_{\rm c}$ furthermore
exhibits sharp dips to zero for every electron added to the QD,
signifying a reversal of the sign of $I_{\rm c}$ due to a $0$-$\pi$
transition of the current-phase relation (see
Fig.~\ref{fig:theory}). All the transition points are seen to
systematically shift toward odd occupation on the QD. In
Fig.~\ref{fig:fit}(a) we have used the phenomenological model of
Ref.~\cite{vecino} with the same parameters as used in
Fig.~\ref{fig:theory} to calculate the full gate-dependence of the
critical current (see Supporting Information, Section~S2). Even
though the QD in reality has an unpolarized magnetic moment in the
odd valleys, the theory, which is based on a polarized
magnetization, does capture two important consequences of the
magnetic moment: \textit{(i)} the magnitude of the critical current
and thus its strong suppression as compared to a non-interacting
resonant level ($I_0=e\Delta/\hbar\simeq 25$\,nA), which is a result
of the transport of Cooper-pairs through a strongly correlated
electronic system. \textit{(ii)} the slight shift of the transition
points toward odd occupation.
\newline\indent
We end by noting that Josephson junctions fabricated in Al-AlO$_{\rm
x}$-Al are already used in superconducting quantum bits (qubits)
\cite{nakamura,vion,chiorescu}. The realization of a well-controlled
Josephson QD junction in a SWCNT, opens the route towards new
superconducting qubits where better control of the Josephson
junctions can be achieved.
\footnotesize{
\subsection{Acknowledgements} We acknowledge fruitful
discussions with M.-S.~Choi, M.~H.~Devoret, T.~Heikkila, V.~Meden,
J.~Paaske, C.~Urbina, and W.~Wernsdorfer. We thank F.~B.~Rasmussen
for technical help. The work was in part supported by the EU-STREP
ULTRA-1D and CARDEQ program, and (T.~N.) by the research plan MSM
0021620834 financed by the Ministry of Education of the Czech
Republic.
\subsection{Supporting information available}
The three sections in the Supporting Information include: (i) To
demonstrate reproducibility we show similar measurements on another
device. (ii) Justification of the fitting procedure and explanation
of the simple model prediction for the supercurrent. (iii)
Estimating the asymmetry and level broadening of the device from the
Coulomb blockade peaks. This material is available free of charge
via the Internet at http://pubs.acs.org.
\end{document}